\documentclass[aps,amsmath,amssymb,nofootinbib]{revtex4}
\usepackage{graphicx}				% Use pdf, png, jpg, or eps§ with pdflatex; use eps in DVI mode
\usepackage{epsfig}
\usepackage{bm}
\usepackage{amssymb}
\usepackage{amsmath}
\usepackage{color}			% TeX will automatically convert eps --> pdf in pdflatex		
\usepackage{amssymb}
\usepackage{wrapfig}

\allowdisplaybreaks
\usepackage{hyperref}
\hypersetup{
  colorlinks   = true,    % Colours links instead of ugly boxes
  urlcolor     = blue,    % Colour for external hyperlinks
  linkcolor    = blue,    % Colour of internal links
  citecolor    = red      % Colour of citations
}

\newcommand{\be}{\begin{equation}}
\newcommand{\ee}{\end{equation}}
\newcommand{\bea}{\begin{eqnarray}}
\newcommand{\eea}{\end{eqnarray}}

\newcommand{\nn}{\nonumber}
\newcommand{\tr}{\mathrm{tr\, }}
\newcommand{\ud}{\, \mathrm{d}}

\def\simge{\mathrel{%
   \rlap{\raise 0.511ex \hbox{$>$}}{\lower 0.511ex \hbox{$\sim$}}}}
\def\simle{\mathrel{
   \rlap{\raise 0.511ex \hbox{$<$}}{\lower 0.511ex \hbox{$\sim$}}}}

\begin{document}
\title{An incorrect pre-asymptotic RG flow of
  scattering amplitudes in QCD towards the unitarity limit }

\author{Adrian Dumitru}
\email{adrian.dumitru@baruch.cuny.edu}
\affiliation{Department of Natural Sciences, Baruch College,
CUNY, 17 Lexington Avenue, New York, NY 10010, USA}
\affiliation{The Graduate School and University Center, The City
  University of New York, 365 Fifth Avenue, New York, NY 10016, USA}

\author{Vladimir Skokov}
\email{vskokov@ncsu.edu}
\affiliation{Department of Physics, North Carolina State University,
  Raleigh, NC 27695}
\affiliation{RIKEN BNL Research Center, Brookhaven National
  Laboratory, Upton, NY 11973, USA}

\date{\today}

\begin{abstract}
Scattering amplitudes in QCD exhibit a definite RG flow with energy
towards the unitarity limit. In this paper we put forward an evolution
equation which allows one to modify continuously the pre-asymptotic RG
flow towards ``saturation'' of Wilson line correlators.  It preserves
the linearly unstable zero field fixed point and the unitarity limit
attractor. We present the evolution equations in a form suitable for
efficient numerical solution.  Hence, the proposed evolution equation
in principle permits phenomenological comparisons of our incorrect
pre-asymptotic RG flows to the BK/JIMWLK flow of QCD. Ultimately, it
may lead to observational constraints on the approach to unitarity,
constraining it to be specifically the one of QCD. However, it appears
that single-inclusive particle spectra in the forward region of p+A
collisions at RHIC alone do not provide stringent constraints on the
evolution.
\end{abstract}

\maketitle

The density of soft gluons in strongly boosted hadrons or nuclei is
non-perturbatively
high~\cite{Gribov:1984tu,Mueller:1989st,Mueller:1999wm,Mueller:2001fv}
thereby saturating the unitarity limit of scattering amplitudes from
such an object.  QCD evolution in rapidity towards the unitarity limit
is described at leading logarithmic (LL) accuracy by the B-JIMWLK
renormalization group equations~\cite{Balitsky:1995ub,
  Balitsky:1998ya, Balitsky:2001re, Jalilian-Marian:1997jx,
  Jalilian-Marian:1997gr, JalilianMarian:1997dw, Iancu:2001ad,
  Iancu:2000hn, Ferreiro:2001qy, Weigert:2000gi}.  These can be cast
in the form of a Langevin process in the space of $V_{\vec
  x}\in\,$SU($N_c$) Wilson
lines~\cite{Weigert:2000gi,Blaizot:2002xy,Rummukainen:2003ns} over the
transverse plane of the collision.

Our approach here is to construct a new RG flow with energy of {\em
  all} correlators of Wilson lines, i.e.\ for the entire Balitsky
hierarchy, directly by modifying the noise correlator of JIMWLK.
Expanding the conventional JIMWLK equation for a small step $\ud y$ in
rapidity and using a noise-noise correlator
\be
\langle \xi_{\vec x}^{ia}\, \xi_{\vec y}^{jb} \rangle = \sigma^2\,
\delta^{ij}\, \delta^{ab}\, \delta_{\vec x \vec y}
\ee
with an arbitrary variance $\sigma^2$, we obtain the modified
Langevin process
\begin{multline}\label{eq:jtimestepexp}
V_{\vec x}(y + \ud y) = V_{\vec x}(y) \bigg[
1
+ i \frac{\sqrt{\alpha_s \ud y }}{\pi}\int_{\vec z} \vec K_{\vec x-\vec z}
\cdot \left( \vec \xi_{\vec z}
- V^\dag_{\vec x}V_{\vec z} \vec \xi_{\vec z} V^\dag_{\vec z} V_{\vec x} \right)
\\
-
\frac{\alpha_s \ud y }{2 \pi^2}
{\sigma^2}
\int_{\vec z} \vec K^2_{\vec x-\vec z}
\left(2 - U^\dag_{\vec x} U_{\vec z}  - U^\dag_{\vec z} U_{\vec x} \right)^{ab} t^a t^b
+
\frac{\alpha_s \ud y }{2 \pi^2}
\int_{\vec z} \vec K^2_{\vec x-\vec z}
\left(U^\dag_{\vec x} U_{\vec z}  - U^\dag_{\vec z} U_{\vec x} \right)^{ab} t^a t^b
\bigg]~.
\end{multline}
We can rewrite adjoint Wilson lines in terms of fundamental ones,
$U_{\vec x}^{ca} t^a_{ij}  = (V_{\vec x}^\dagger t^c V_{\vec x} )_{ij}$ to
obtain the
equivalent form
\begin{eqnarray}
V_{\vec x}(y + \ud y) &=& V_{\vec x}(y) \bigg[
1 + i \frac{\sqrt{\alpha_s \ud y }}{\pi}\int_{\vec z} \vec K_{\vec
  x-\vec z} \cdot \left(
\vec \xi_{\vec z} - V^\dag_{\vec x} V_{\vec z} \vec \xi_{\vec z} V^\dag_{\vec z} V_{\vec x} \right)
\nn\\
& & -
\frac{\alpha_s \ud y }{2 \pi^2}
\int_{\vec z} \vec K^2_{\vec x-\vec z}
\left(  \sigma^2 N_c      -
\sigma^2 V^\dagger_{\vec x} V_{\vec z}    \tr  (V_{\vec x} V_{\vec z}^\dagger)
+\frac{ \sigma^2 -1 }{2}  V^\dagger_{\vec x} V_{\vec z}
\tr  (V_{\vec x} V_{\vec z}^\dagger)
-
\frac{ \sigma^2 -1 }{2}  V^\dagger_{\vec z} V_{\vec x}
\tr  (V_{\vec z} V_{\vec x}^\dagger)
\right )
\bigg]~.
\label{eq:OneTimeStep_lin}
\end{eqnarray}
Here, $\vec K_{\vec x} = \vec x/x^2$ is the ``square root'' of the
BFKL kernel~\cite{Kuraev:1977fs,Balitsky:1978ic}.  This equation
suffices for the derivation of evolution equations of specific Wilson
line operators. However, for numerical solution the evolution should
be cast in a form that preserves $V_{\vec x} \in \,$SU($N_c$)
exactly. With some algebra one can show that this is achieved by
writing
\begin{eqnarray}
V_{\vec x}(y + \ud y) &=&
\exp\left\{-i \sigma^{-2} \frac{\sqrt{\alpha_s \ud y }}{\pi}\int_{\vec z}
  \vec K_{\vec x-\vec z}   \cdot ( V_{\vec z} \vec\xi_{\vec z} V^\dagger_{\vec z}) \right\}
\nn\\
& & \times
V_{\vec x}
\exp\left\{i\frac{\sqrt{\alpha_s \ud y }}{\pi}\int_{\vec z}
  \vec K_{\vec x-\vec z} \cdot
  \left[ \vec \xi_{\vec z}  - (1-\sigma^{-2})  V_{\vec x}^\dagger
    V_{\vec z} \vec \xi_{\vec z}
    V_{\vec z}^\dagger V_{\vec x}     \right]   \right\}~.
\label{eq:OneTimeStep_exp}
\end{eqnarray}
This reproduces eq.~(\ref{eq:OneTimeStep_lin}) to linear order in $\ud
y$. Also, just like in standard JIMWLK evolution the implementation of
eq.~(\ref{eq:OneTimeStep_exp}) requires just two (matrix valued) FFTs
per rapidity step to compute the arguments of the exponentials for all
$\vec x$ (which is crucial for numerical solutions to be feasible).
In the limit, $\sigma \to 1$, eq.~\eqref{eq:OneTimeStep_exp} 
reproduces LL JIMWLK evolution.

From eq.~(\ref{eq:OneTimeStep_lin}) we can derive the evolution
equation for the expectation value of the dipole scattering matrix
$S_{\vec x \vec y} = \tr V_{\vec x} V^\dagger_{\vec y} / N_c$:
\bea
\frac{\partial}{\partial Y} \left< S_{\vec x \vec y} \right> &=&
\frac{\bar\alpha_s\, \sigma^2}{2\pi} \int_{\vec z}\left\{
\frac{(\vec x - \vec y)^2}{(\vec x - \vec z)^2\, (\vec y - \vec z)^2}
\left[- \left< S_{\vec x \vec y} \right> +
  \left< S_{\vec x \vec z}\, S_{\vec z \vec y} \right> \right] \right.\nn\\
& & \left. + \frac{1}{(\vec x - \vec z)^2}\frac{1-\sigma^2}{2\sigma^2}
\left[
   \left< S_{\vec x \vec z}\, S_{\vec z
    \vec y} \right>
  - \left< S_{\vec z \vec x}\, Q_{\vec x \vec z \vec x \vec y}\right>
  \right]
+
\frac{1}{(\vec y - \vec z)^2}\frac{1-\sigma^2}{2\sigma^2}
\left[
   \left< S_{\vec x \vec z}\, S_{\vec z
    \vec y} \right>
  - \left< S_{\vec y \vec z}\, Q_{\vec x \vec y \vec z \vec y}\right>
  \right]
\right\}~,
\label{eq:Sxy-evol}
\eea
where $\bar\alpha_s = \alpha_s N_c/\pi$.  For $\sigma^2=1$ this, of
course, reduces to the B-JIMWLK evolution equation for the
dipole. However, for $\sigma^2\ne1$ the r.h.s.\ involves the
quadrupole $Q_{\vec x \vec y \vec u \vec v} = \tr V_{\vec x}
V^\dagger_{\vec y} V_{\vec u} V^\dagger_{\vec v}/ N_c$. Note that UV
divergences for $\vec z \to \vec x, \vec y$ cancel for any value of
$\sigma^2$. Diagrammatically~\cite{Mueller:2001uk,Kovchegov:2012mbw},
the modification to the dipole evolution equation is due to a factor
of $\frac{1+\sigma^2}{2\sigma^2}$ for the ``real emission'' diagrams
where one of the quarks emits and reabsorbs a gluon; and due to a new
contribution from this diagram corresponding to dipole $\rightarrow$
dipole + quadrupole splitting. On the other hand, the virtual
corrections and the real emission diagrams where the quark and the
anti-quark exchange a gluon are unmodified. Therefore, we conjecture
that no theory (defined in terms of Feynman diagrams) exists which
corresponds to the evolution equation~(\ref{eq:Sxy-evol}) for the
dipole, when $\sigma^2 \ne 1$.\\

Equation (\ref{eq:Sxy-evol}) has two fixed points: one corresponds to
the zero field limit $S,Q=1$, the other to the strong field/unitarity
limit $S=Q=1$. We require the former to be linearly unstable,
and the latter to be attractive. This restricts $\sigma^2 >
\frac{1}{2}$. In the limit $\sigma^2\to\infty$, for example, the
evolution of $\langle S_{\vec x \vec y} \rangle$ with rapidity is
described by
\bea
\frac{\partial}{\partial Y} \left< S_{\vec x \vec y} \right> &=&
\frac{\bar\alpha_s \,\sigma^2}{2\pi} \int_{\vec z}\left\{
\frac{(\vec x - \vec y)^2}{(\vec x - \vec z)^2\, (\vec y - \vec z)^2}
\left[- \left< S_{\vec x \vec y} \right> +
  \left< S_{\vec x \vec z}\, S_{\vec z \vec y} \right> \right] \right.\nn\\
& & \left. - \frac{1}{2(\vec x - \vec z)^2}
\left[
   \left< S_{\vec x \vec z}\, S_{\vec z
    \vec y} \right>
  - \left< S_{\vec z \vec x}\, Q_{\vec x \vec z \vec x \vec y}\right>
  \right]
-
\frac{1}{2(\vec y - \vec z)^2}
\left[
   \left< S_{\vec x \vec z}\, S_{\vec z
    \vec y} \right>
  - \left< S_{\vec y \vec z}\, Q_{\vec x \vec y \vec z \vec y}\right>
  \right]
\right\}~,
\label{eq:Sxy-evol-sigma-00}
\eea
which involves the effective coupling $\bar\alpha_s \,\sigma^2$.

\begin{figure}
\centering
%\begin{wrapfigure}{R}{0.5\textwidth}
%\centering
\includegraphics[width=0.5\textwidth]{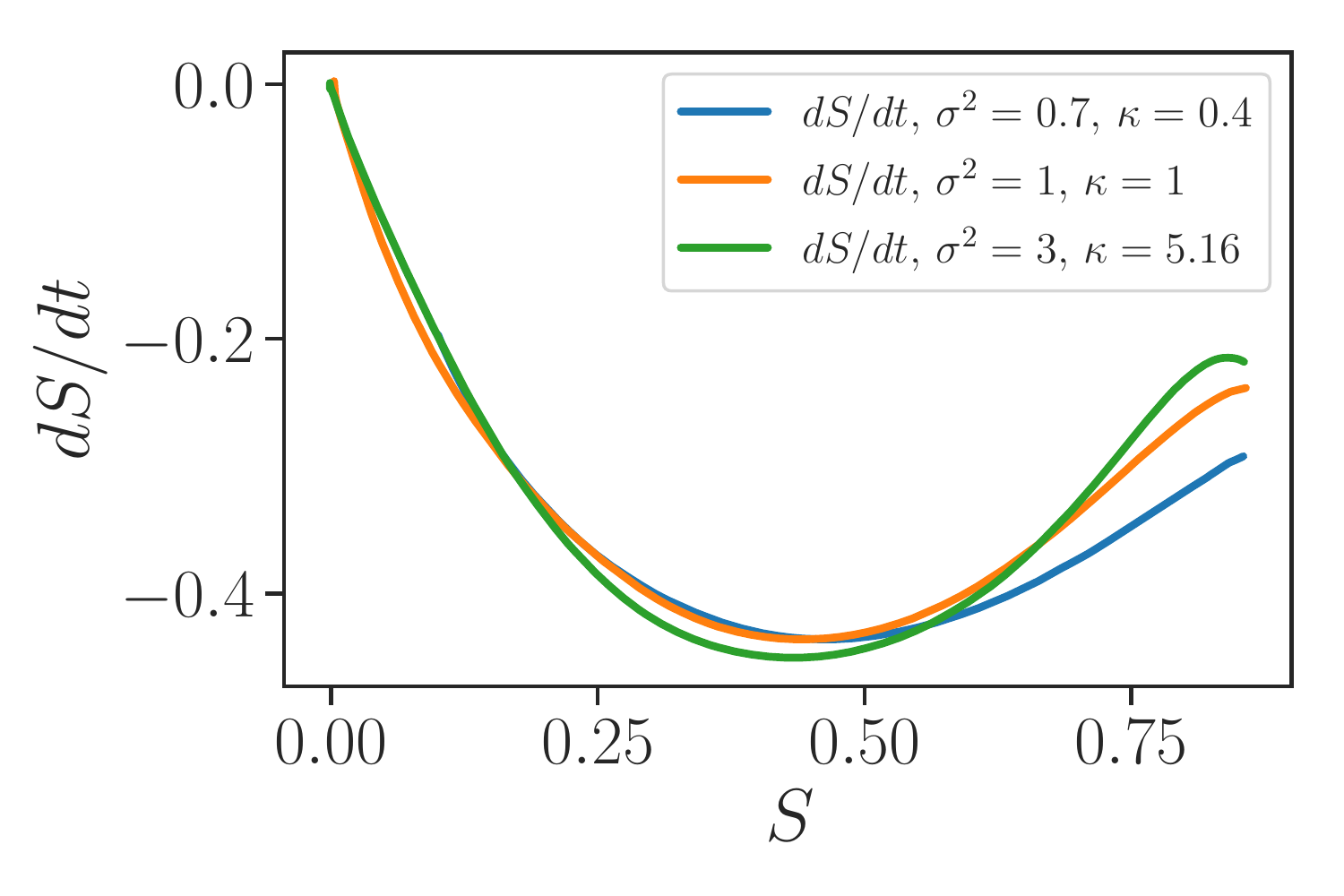}
\caption{\label{fig:Splot} Evolution trajectory of the averaged dipole
  $S$-matrix at fixed $\vec x - \vec y$ for BK-JIMWLK ($\sigma^2=1$)
  vs.\ modified flow ($\sigma^2=0.7, 3$). These curves are numerical
  solutions of eq.~(\ref{eq:OneTimeStep_exp}); $t = \bar\alpha_s Y$
  denotes the evolution time. For the runs with $\sigma^2\ne 1$ the coupling
  constant was rescaled to $\bar\alpha_s \to \bar\alpha_s / \kappa$.
  \label{fig:SdS}}
%\end{wrapfigure}
\end{figure}
Applying a large-$N_c$ mean field
approximation~\cite{Kovchegov:1999yj, Kovchegov:1999ua} $\left< S\,
S\right> \to \left< S\rangle\, \langle S\right>$, $\left< S\, Q\right>
\to \left< S\rangle\, \langle Q\right>$, and assuming $\langle
Q\rangle \sim \langle S\rangle^2$~\footnote{This does not refer to a
  naive factorization $\langle Q_{\vec x \vec y \vec u \vec v}\rangle
  \to \langle S_{\vec x \vec y}\rangle \, \langle S_{\vec u \vec
    v}\rangle$. Rather, here we use that in the approach to unitarity
  the typical magnitude of the quadrupole is of order the squared
  magnitude of the dipole. See refs.~\cite{Iancu:2011ns,
    Dominguez:2011wm} for detailed discussions.}, we find that the
zero field fixed point is linearly unstable just like for standard BK
evolution~\cite{Balitsky:1995ub, Balitsky:1998ya, Balitsky:2001re,
  Kovchegov:1999yj, Kovchegov:1999ua} (however, the eigenvalue is
different). On the other hand, the unitarity limit is asymptotically
(linearly) stable and attractive. However, while the BK/JIMWLK flow to
the $\langle S\rangle=0$ fixed point exhibits ``repulsion'' at
quadratic order in $\langle S\rangle$, in
eq.~(\ref{eq:Sxy-evol-sigma-00}) the repulsion is pushed to cubic
order. Figure~\ref{fig:SdS} compares the evolution of the dipole
$S$-matrix at fixed $\vec r = \vec x - \vec y$ for BK-JIMWLK vs.\ our
modified flow, starting from MV
model~\cite{McLerran:1994ni,McLerran:1994ka,Lappi:2007ku} initial
conditions where $\langle S\rangle$ is real.  We have rescaled the
coupling constant, which is a free parameter in the LL evolution
equation, to approximately match the two curves. Hence, increasing
$\sigma^2$ from its value $\sigma^2=1$ in QCD one is able to modify
the pre-asymptotic RG flow to the fixed point corresponding to the
unitarity limit.

It is also interesting to note the correlated contribution to dipole pair
evolution,
\bea
& &
\frac{\partial}{\partial Y} \left< S_{\vec x \vec y}\, S_{\vec r \vec s} \right> -
\left< S_{\vec x \vec y}\right>\, \frac{\partial}{\partial Y}
\left<S_{\vec r \vec s}\right> -
\left<S_{\vec r \vec s}\right>\, \frac{\partial}{\partial Y}\left<
S_{\vec x \vec y}\right>
=\nn\\& &\quad - \frac{\bar\alpha_s\, \sigma^2}{2\pi\, N_c}
\int_{\vec z} \left(\vec K_{\vec x - \vec z} -\vec K_{\vec y - \vec z}\right)\cdot
\left(\vec K_{\vec r - \vec z} -\vec K_{\vec s - \vec z}\right)
\left[Q_{\vec x \vec s \vec r \vec y} +
  Q_{\vec r \vec s \vec x \vec y} -
  {\cal S}_{\vec z \vec s \vec r \vec z \vec x \vec y} -
  {\cal S}_{\vec z \vec y \vec x \vec z \vec r \vec s} \right]~,
\eea
where ${\cal S}_{\vec z \vec s \vec r \vec z \vec x \vec y} = \tr
V_{\vec z} V^\dagger_{\vec s} V_{\vec r} V^\dagger_{\vec z} V_{\vec x}
V^\dagger_{\vec y}\, /N_c$ is a trace over six Wilson lines. The
r.h.s.\ does not involve any new operators as compared to the JIMWLK
equation; it has just been rescaled to the effective coupling
$\bar\alpha_s\, \sigma^2$. This is due to the fact that this
contribution arises from the second term in
eq.~(\ref{eq:OneTimeStep_lin}), i.e.\ the one linear in the noise
$\xi^{ia}_{\vec x}$, which has the same structure as in JIMWLK
evolution. Therefore, the correlated contribution to dipole pair
scattering is affected differently by our modification of the flow
than the evolution of the $S$-matrix of a single dipole. Two-particle
or dijet correlations~\cite{Albacete:2010pg, Kovner:2011pe,Lappi:2012nh,
  Dumitru:2014yza, Shi:2020djm, Stasto:2011ru,
  Zheng:2014vka, Albacete:2018ruq, Stasto:2018rci, Kolbe:2020tlq}
could provide  means to constrain evolution at small-$x$.\\

\begin{figure}
\centering
\includegraphics[width=0.65\textwidth]{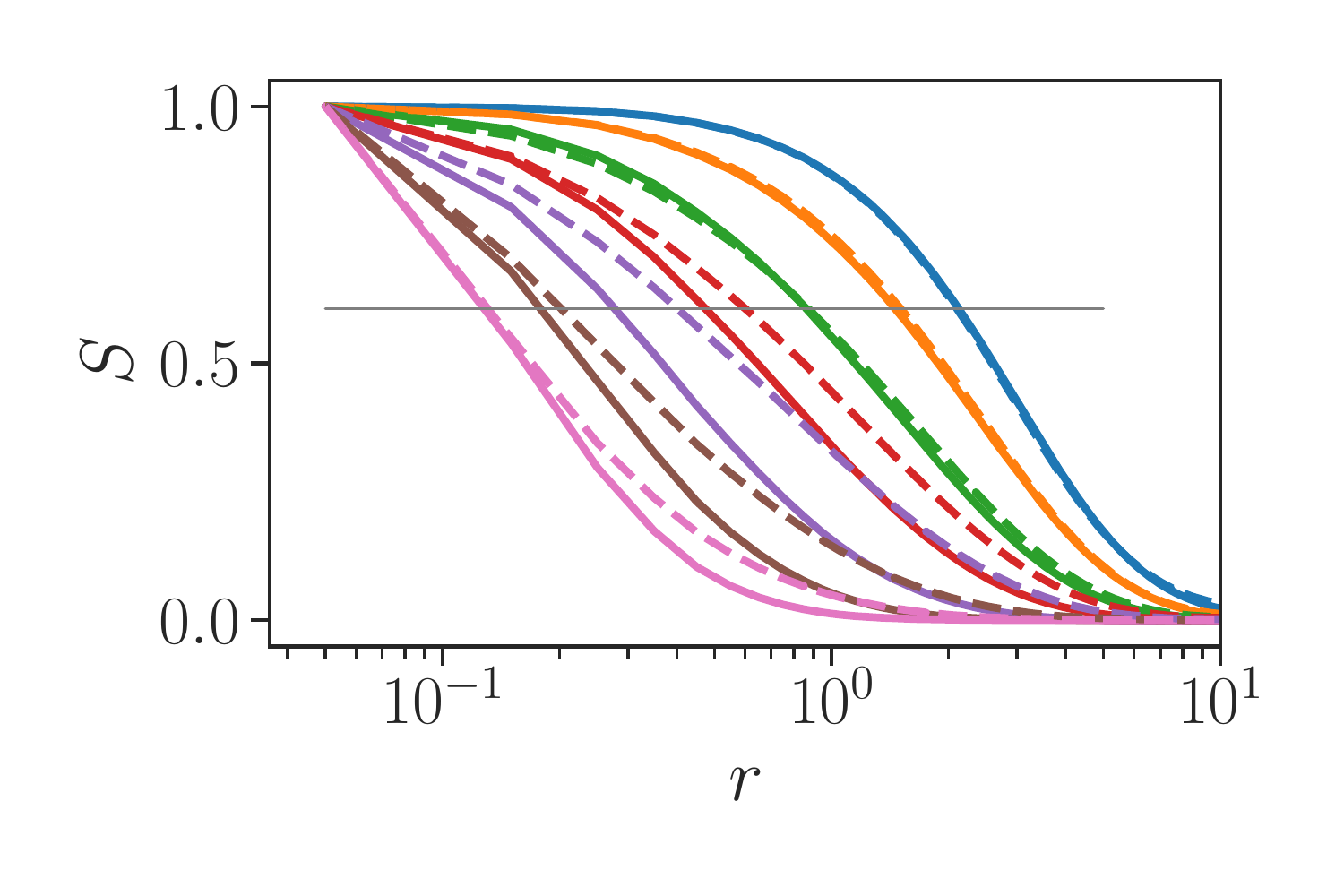}
\caption{The average dipole $S$-matrix as a function of dipole size
  $r$ at different evolution times $t=\bar\alpha_s Y=0, \dots,
  6$. Solid lines correspond to BK-JIMWLK ($\sigma^2=1$) evolution
  while dashed lines correspond to modified flow at $\sigma^2=3$ (here
  with the coupling rescaled as $\bar\alpha_s \to \bar\alpha_s /
  6$). These curves are numerical solutions of
  eq.~(\ref{eq:OneTimeStep_exp}). The horizontal line at
  $S(r)=1-1/e\simeq0.63$ indicates the transition to the non-linear
  regime. Units for $r$ are given by the RMS color charge density
  $g^2\mu$ of the MV model.
  \label{fig:wave}}
\end{figure}
For $\sigma^2=3$ we have checked the ``travelling wave''
solutions~\cite{Munier:2003vc,Munier:2003sj} for $\langle S_{\vec x
  \vec y}\rangle(Y)$ which lead to geometric scaling of the cross
section for scattering of a virtual photon from a
proton~\cite{Stasto:2000er}. In fig.~\ref{fig:wave} we have again
rescaled the coupling constant ($\bar \alpha_s \to
\bar\alpha_s/\kappa$) so that the speed at $S=1-1/e$ is similar to
that for BK-JIMWLK evolution. We do observe a modification of the
shape of the travelling waves at intermediate rapidities, deep in the
non-linear regime at small $S$.  For $\sigma^2<1$, the rescaling
factor $\kappa$ is $<1$. In particular, we confirmed numerically that
the evolution of $\langle S_{\vec x \vec y}\rangle$ ``stalls''
($\kappa\to0$) for a value of $\sigma^2$ slightly less than 1/2.\\

The modified evolution equation~(\ref{eq:OneTimeStep_lin},
\ref{eq:OneTimeStep_exp}) permits phenomenological comparisons of our
incorrect evolution (towards unitarity) to the BK/JIMWLK flow of
QCD. Ultimately, this would provide observational constraints on the
RG flow (bounds on $|1-\sigma^2|$), i.e.\ on the approach towards
``saturation'' of QCD scattering amplitudes. The question we ask is
not whether or not unitarity is achieved. Rather, we are interested in
constraining from observation the flow towards unitarity to be the
one of QCD.\\

\begin{figure}
\centering
\includegraphics[width=0.65\textwidth]{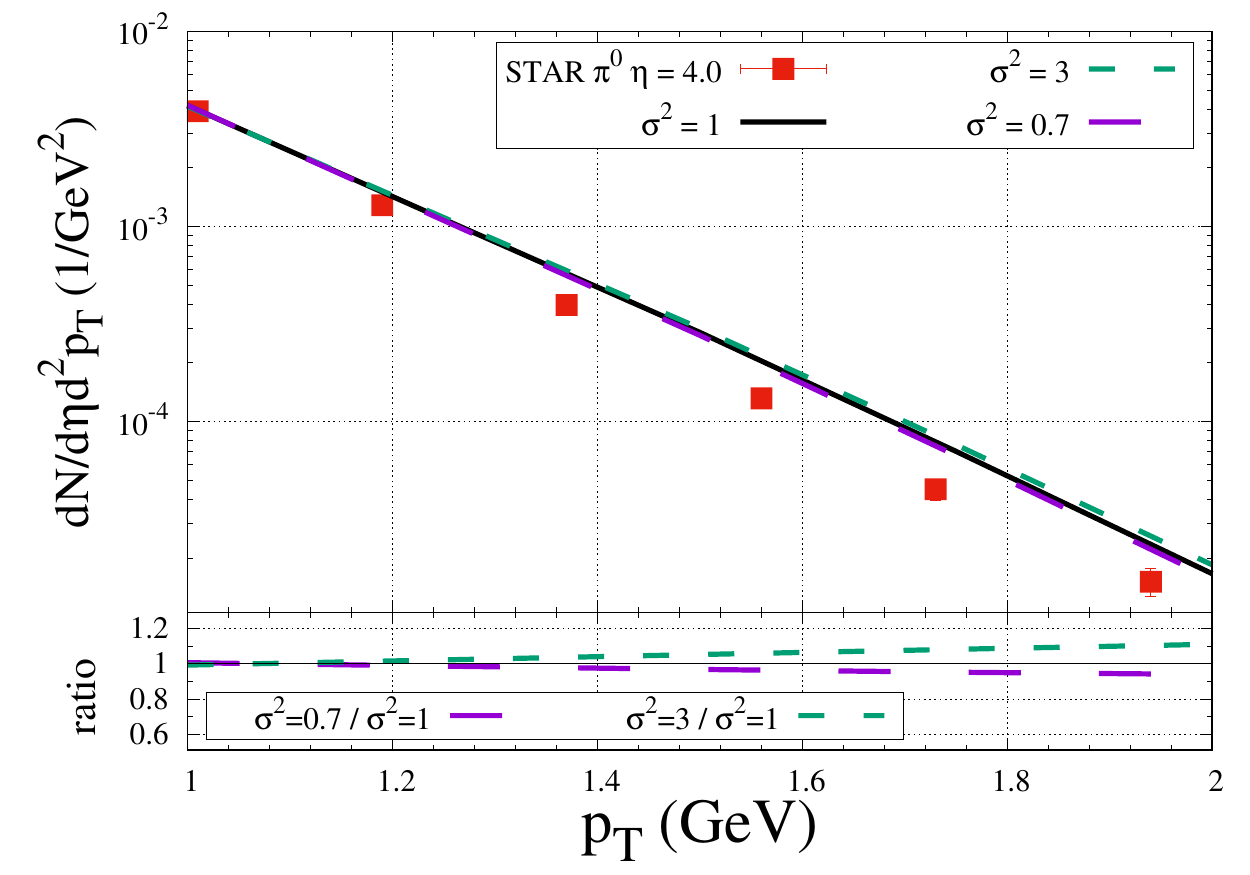}
\caption{Single-inclusive $\pi^0$ transverse momentum
  distribution in $p+Au$ collisions at $\sqrt{s}=200$~GeV and rapidity
  $\eta=4$. We compare model calculations (see text) based on
  small-$x$ evolution at various values of $\sigma^2$, where
  $\sigma^2=1$ is the LL JIMWLK evolution of QCD. For reference we
  also show data by the STAR collaboration~\cite{Adams:2006uz}.
  \label{fig:forward-pA}}
\end{figure}
In fig.\ref{fig:forward-pA} we compare single-inclusive
$p_T$-distributions obtained from fixed-coupling ($\alpha_s=0.1$)
evolution at various values for $\sigma^2$; we employ the ``hybrid
formalism'' of ref.~\cite{Dumitru:2005gt} to compute the $p_T$
spectrum. Realistic phenomenology of QCD evolution requires one to
account for the running of the
coupling~\cite{Albacete:2007sm,Albacete:2007yr}. Its implementation in
B-JIMWLK evolution has been discussed in
ref.~\cite{Lappi:2012vw}\footnote{The running coupling prescription
  suggested in ref.~\cite{Lappi:2012vw} also introduces a quadrupole
  operator in the evolution of the dipole. However, those are specific
  NLO corrections associated with running ($\beta(\alpha_s)\sim -
  \alpha_s^2$) of $\alpha_s$. They do not correspond to a modification
  of the proper QCD evolution equation like in our scenario.}, and
references therein. One should also account for NLO corrections to
particle production~\cite{Altinoluk:2011qy, Chirilli:2011km,
  Stasto:2013cha, Ducloue:2017dit}.  However, our main goal here is
not to obtain a good fit to the data but rather to check the
sensitivity of the spectra to $\sigma^2$. As $\sigma^2$ was varied we
performed no adjustment of the coupling in the evolution equation,
$\alpha_s=0.1$, of the initial saturation scale, $Q_s(0)=1$~GeV on
average over impact parameters, of the {\em form} of the initial
ensemble of Wilson lines at $Y=0$ (MV
model~\cite{McLerran:1994ni,McLerran:1994ka,Lappi:2007ku}), or of the
scales for the proton parton distribution functions~\cite{Lai:1999wy}
and the fragmentation functions~\cite{Kniehl:2000fe}, $Q^2=p_T^2$.
The figure shows that the single-inclusive spectra in the forward
region of pA collisions at RHIC {\em alone} do not constrain the
evolution to be close to that for QCD.\\

The modification of the approach to the asymptotic fixed point should
affect suitable observables. The wealth of data from HERA, RHIC,
LHC, and in the future from the EIC, may make it possible to set
quantitative constraints on the flow towards the asymptotic limit of
QCD.  A challenge for small-$x$ QCD phenomenology is to provide a
bound $|1-\sigma^2|/\sigma^2 \ll 1$.

\begin{acknowledgments}
A.D.\ acknowledges support by the DOE Office of Nuclear Physics
through Grant No.\ DE-FG02-09ER41620; and from The City University of
New York through the PSC-CUNY Research grant 62098-00~50.

V.S. acknowledges
support by the DOE Office of Nuclear Physics through
Grant No. DE-SC0020081. V.S. thanks the ExtreMe Matter Institute EMMI (GSI
Helmholtzzentrum f\"ur Schwerionenforschung, Darmstadt,
Germany) for partial support and hospitality. 

\end{acknowledgments}

\begin{appendix}
\section{Reweighting}

Here we outline briefly how operator expectation values in the
standard B-JIMWLK ensemble can be reconstructed from a simulation with
$\sigma^2\ne 1$ through reweighting.

Consider an operator $O$ given by products of Wilson lines at
different transverse points. Upon taking a step in rapidity its
expectation value in the standard B-JIMWLK ensemble ($\sigma^2= 1$)
evolves as follows:
\be
\langle O\rangle(Y+\ud y) = \langle O\rangle(Y) +
\ud y \, \langle O'\rangle(Y) ~. 
\ee
The operator $O'$ is obtained from $O$ by replacing all Wilson lines
in $O$ by the r.h.s.\ of eq.~(\ref{eq:OneTimeStep_lin}), setting
$\sigma^2=1$; this is followed by an expansion in $\ud y$ to
linear order.

The brackets $\langle\cdot\rangle$ indicate an average over the noise
(in the rapidity slice $Y$):
\be
\langle O'\rangle = \left[ \prod\limits_{\vec x, a, i}
  \int \ud \xi_{\vec x}^{ia}\,
  P_{\sigma^2=1}(\xi_{\vec x}^{ia})
  \right] \, O'(\left\{\xi_{\vec x}^{ia}  \right\}) ~~~~;~~~~
P_{\sigma^2}(\xi_{\vec x}^{ia}) = \frac{1}{\sqrt{2\pi \sigma^2}}\,
e^{- {\xi_{\vec x}^{ia} \xi_{\vec x}^{ia}}/{2\sigma^2}}~.
\label{eq:<O'>_sigma2=1}
\ee
With $w(\xi_{\vec x}^{ia}) = P_{\sigma^2=1}(\xi_{\vec x}^{ia}) /
P_{\sigma^2}(\xi_{\vec x}^{ia})$ we can also write this in the form
\be
\langle O'\rangle = \left[ \prod\limits_{\vec x, a, i}
  \int \ud \xi_{\vec x}^{ia}\, w(\xi_{\vec x}^{ia})
  P_{\sigma^2}(\xi_{\vec x}^{ia})  \right] \, O'(\left\{\xi_{\vec x}^{ia}  \right\})
\equiv \langle O' \, w\rangle_{\sigma^2}
\label{eq:<O'>_sigma2/=1}
~.
\ee
Hence, $\langle O'\rangle$ can be computed by averaging over noise
with variance $\sigma^2\ne1$ provided that one multiplies $O'$ by the
weight $w = \prod_{\vec x, a, i}\, w(\xi_{\vec x}^{ia})$. However, we
repeat that $O'$ must be obtained via the r.h.s.\ of
eq.~(\ref{eq:OneTimeStep_lin}) with $\sigma^2=1$.\\

In general it is not possible to compute the averages over the noise
exactly. Rather, one must employ Monte-Carlo importance
sampling. Eq.~(\ref{eq:<O'>_sigma2=1}) will then lead to much more
accurate results than eq.~(\ref{eq:<O'>_sigma2/=1}) since the weight
$w$ will fluctuate strongly from configuration to configuration [unless
$|\sigma^2 -1 | \simle (A_\perp d_A)^{-1/2}$, where $d_A=N_c^2-1$ and
$A_\perp$ is the area, i.e.\ the number of sites, of the transverse lattice].\\

Nevertheless, reweighted averages via eq.~(\ref{eq:<O'>_sigma2/=1})
can be useful for some applications. For example, consider computing
the expectation value of $O'$ over a {\em biased} ensemble, where one
is interested in selecting rare evolution trajectories. In the
standard B-JIMWLK ensemble this corresponds to evaluating $\langle
O'\, b\rangle$ where $b$ denotes a bias such as high gluon
multiplicity or mean transverse momentum, for which a simple model has
been considered in refs.~\cite{Dumitru:2017cwt, Dumitru:2018iko,
  Kapilevich:2019sie}. A simpler example in the present context would
be a bias of the form $b = \prod_{\vec x, a, i}\, \Theta\!\left(\Xi -
|\xi_{\vec x}^{ia}|\right)$. If $\Xi\gg1$ ($\Xi\ll1$) this bias
evidently prefers a wider (narrower) distribution for the noise which
is better sampled with $P_{\sigma^2\sim\Xi}(\xi_{\vec x}^{ia})$ than
with $P_{\sigma^2=1}(\xi_{\vec x}^{ia})$. A more physical example is
\begin{equation}
b = \exp \left\{ \lambda  \frac{\alpha_s^2} { N_c^4 }\frac{1}{A_\perp}
\int \frac{d^2 q}{(2\pi)^2} \left[ A^i_a (\vec q) A^i_a(-\vec q)  -
  \frac{1}{q^2} q^i A^{i}_a (\vec q) \ q^j A^{j}_a(-\vec q) \right]^2
\right\}\,, \label{eq:b_Ai}
\end{equation}
which suppresses longitudinal and enhances transverse light-cone gauge
fields (if $\lambda>0$)
\begin{equation}
  A_i(x) = \frac{1}{ig } V^\dagger_{\vec x} \partial_i V_{\vec x}~.
\end{equation}
Again, if $b$ selects rare evolution trajectories one should expect it
to exhibit very large fluctuations across configurations. This would
lead to large errors for $\langle O'\, b\rangle$ since this
average would be completely dominated by a small subset of
configurations. Computing, instead, evolution trajectories for
$\sigma^2\ne1$, and reweighted averages $\langle O'\, b\,
w\rangle_{\sigma^2}$, could in some cases increase the overlap with
the desired ensemble, i.e.\ if the product $b\, w$ is approximately
constant for a greater set of configurations. For some biases $b$ like
the one in eq.~(\ref{eq:b_Ai}) it would be beneficial to employ
correlated (in $\vec x$) and non-diagonal (in color) noise.

\end{appendix}

\bibliography{spires}

\end{document}